\documentstyle[aps,epsf,bezier,prl%
,twocolumn%
]{revtex}
\begin{document}
\draft
\title{Nonlinear Dynamics of Dry Friction}
\author{Franz-Josef Elmer}
\address{Institut f\"ur Physik, Universit\"at
   Basel, CH-4056 Basel, Switzerland}
\date{appears in Journal of Physics A}
\maketitle
\begin{abstract}
The dynamical behavior caused by dry friction is studied for a 
spring-block system pulled with constant velocity over a surface.
The dynamical consequences of a general type of phenomenological
friction law (stick-time dependent static friction, velocity
dependent kinetic friction) are investigated.  Three types of motion
are possible:  Stick-slip motion, continuous sliding, and
oscillations without sticking events. A rather complete discussion of
local and global bifurcation scenarios of these attractors and their
unstable counterparts is present.
\end{abstract}
\pacs{PACS numbers: 03.20.+i, 46.30.Pa, 81.40.Pq}

\narrowtext

Since more than 200~years Coulomb's laws of dry friction have been
well-known \cite{bow.54}. They state that the friction force is given
by a material parameter (friction coefficient) times the normal
force. The coefficient of static friction (i.e., the force necessary
to start sliding) is always equal to or larger than the coefficient
of kinetic friction (i.e., the force necessary to keep sliding at a
constant velocity).

The dynamical behavior of a mechanical system with dry friction is 
nonlinear because Coulomb's laws distinguish between static friction 
and kinetic friction. If the kinetic friction coefficient is less
than the static one {\em stick-slip motion\/} occurs where the
sliding surfaces alternately switch between sticking and slipping in
a more or less regular fashion \cite{bow.54}. This jerky motion leads
to the everyday experience of squeaking doors and singing violins.

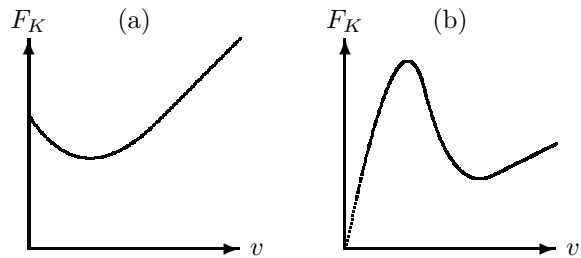
\begin{figure}
\unitlength=0.7mm
\begin{picture}(120,50)
\thicklines
\put(5,5){\vector(1,0){40}}
\put(47,5){\makebox(0,0)[l]{$v$}}
\put(5,5){\vector(0,1){40}}
\put(5,47){\makebox(0,0)[b]{$F_K$}}
\put(25,47){\makebox(0,0)[b]{(a)}}
\bezier{90}(5,30)(15,15)(29,29)
\bezier{60}(29,29)(37,37)(45,45)
\put(65,5){\vector(1,0){40}}
\put(107,5){\makebox(0,0)[l]{$v$}}
\put(65,5){\vector(0,1){40}}
\put(65,47){\makebox(0,0)[b]{$F_K$}}
\put(85,47){\makebox(0,0)[b]{(b)}}
\bezier{120}(65,5)(75,55)(80,35)
\bezier{90}(80,35)(85,15)(93,19)
\bezier{70}(93,19)(99,22)(105,25)
\end{picture}
\caption[ffk]{\protect\label{f.fk}Schematical sketches of typical
velocity dependent kinetic friction laws for systems (a) without and
(b) with boundary lubrication.}
\end{figure}

Even though Coulomb's laws are simple and well-established (many
calculations in engineering rely on these laws), they cannot be
derived in a rigorous way because dry friction is a process which
operates mostly far from equilibrium.  It is therefore no surprise
that deviations from Coulomb's laws have often been found in
experiments. Typical deviations are the following: (i) Static
friction is not constant but increases with the sticking time
\cite{rab.65,hes.94}, i.e., the time since the two sliding surfaces
have been in contact without any relative motion. (ii) Kinetic
friction depends on the sliding velocity; for very large velocities,
it increases roughly linearly with the sliding velocity like in
viscous friction. Coming from large velocities, the friction first
decreases, goes through a minimum, and then increases
\cite{hes.94,bur.67}. In the case of boundary lubrication (i.e., a
few monolayers of some lubricant are between the sliding surfaces) it
decreases again for very low velocities (see Fig.~\ref{f.fk})
\cite{bhu.95,ber.96}. The coefficient of kinetic friction as a
function of the sliding velocity has therefore at least one
extremum.  The kinetic friction can exceed the static friction, but
in the limit of zero sliding velocity it is still less than or equal
to the static friction.

The aim of this paper is to give a rather complete discussion of the
nonlinear dynamics of a single degree of freedom for an {\em arbitrary
phenomenological\/} dry friction law in the sense mentioned above.
This goes beyond the discussion of specific laws found in the
literature \cite{bow.54,hes.94,bur.67,ber.96,per.94,per.95,vet.90}. A
phenomenological law for the friction force depends only on the
macroscopic degrees of freedom. This implies that all microscopic
degrees of freedom are much faster than the macroscopic ones. At the
end of this paper I will give a simple argument why this assumption
will not always be valid. To reveal this invalidity on the macroscopic 
level, it is therefore important to have a {\em complete knowledge
of the dynamical behavior under the assumption that this time-scale
separation works\/}.

There are two other important reasons for knowing the consequences of
the different dry friction laws:  (i) Friction coefficients can be
measured only within an apparatus (e.g. the surface force apparatus
\cite{isr.85} or the friction force microscope \cite{mat.87}). Below
we will see that the dynamical behavior of the whole system is
strongly determined by the friction force and the properties of the
apparatus. For example, stick-slip motion makes it difficult to
obtain directly the coefficient of kinetic friction as a function of
the sliding velocity. Thus, the influence of the measuring apparatus
cannot be eliminated. (ii) Dry friction plays also an important role
in granular materials \cite{jae.96}. An open question there is
whether or not the cooperative behavior of many interacting grains is
significantly influenced by the dynamical behavior due to
modifications of Coulomb's laws.

\begin{figure}
\unitlength=1mm
\begin{picture}(80,40)
\thicklines
\put(15,10){\line(1,0){60}}
\multiput(17,10)(2,0){30}{\line(-1,-1){4}}
\put(65,2){\line(-1,0){15}}
\put(65,2){\line(-2,1){4}}
\put(65,2){\line(-2,-1){4}}
\put(67,2){\makebox(0,0)[l]{$v_0$}}
\put(5,11){\line(0,1){10}}
\multiput(5,13)(0,2){5}{\line(-1,-1){4}}
\put(5,16){\line(1,0){3}}
\multiput(8,16)(8,0){4}{\line(1,2){2}}
\multiput(10,20)(8,0){4}{\line(1,-2){4}}
\multiput(14,12)(8,0){4}{\line(1,2){2}}
\put(40,16){\line(1,0){3}}
\put(26,22){\makebox(0,0)[b]{$\kappa$}}
\put(43,10.2){\framebox(16,11.8){$M$}}
\put(51,24.4){\line(0,-1){2.4}}
\put(51,28){\line(1,-2){1.8}}
\put(51,28){\line(-1,-2){1.8}}
\put(52.8,24.4){\line(-1,0){3.6}}
\put(54,25){\makebox(0,0)[l]{$x$}}
\put(28,29){\line(1,0){48}}
\multiput(28.5,29)(2,0){24}{\line(0,1){2}}
\multiput(32.5,31)(10,0){5}{\line(0,1){1}}
\put(42.5,34){\makebox(0,0)[b]{0}}
\end{picture}
\caption[fsys]{\protect\label{f.sys}A harmonic oscillator with dry
friction.}
\end{figure}
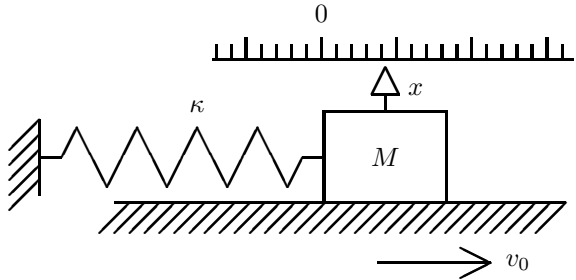

The mechanical environment (e.g. the apparatus) of two sliding
surfaces may have many macroscopic degrees of freedom. The most
important one is the lateral one. Here only systems are discussed
which can be well described by this single degree of freedom. 
Figure~\ref{f.sys} shows the apparatus schematically. It is described
by a harmonic oscillator where a block (mass $M$) is connected via a
spring (stiffness $\kappa$) to a fixed support (see
Fig.~\ref{f.sys}). The block is in contact with a surface which
slides with constant velocity $v_0$. The interaction between the
block and the sliding surface is described by a sticking-time
dependent static friction force $F_S(t_{stick})$ and a velocity
dependent kinetic friction force $F_K(v)$. For the equation of motion
we have to distinguish whether the block sticks or slips. If it
sticks its position $x$ grows linearly in time until the force in the
spring (i.e., $\kappa x$) exceeds the static friction $F_S$. Thus
\begin{mathletters}\label{eqm}
\begin{equation}
  \dot x=v_0,\quad{\rm if}\ |x|\le F_S(t-t_r)/\kappa,
  \label{eqm.stick}
\end{equation}
where $t_r<t$ is the time at which the block has sticked again after
a previous sliding state. If the block slips the equation of motion
reads
\begin{eqnarray}
  \lefteqn{M\ddot x+\kappa x=\,{\rm sign}\,(v_0-\dot x)
   F_K(|v_0-\dot x|),}\label{eqm.slip}\\&&\hspace{30mm}
   {\rm if}\ \dot x\neq v_0\ {\rm or}\ |x|>F_S(0)/\kappa,\nonumber
\end{eqnarray}
\end{mathletters}
where ${\rm sign}\,(x)$ denotes the sign of $x$.

We start our investigation with Coulomb's laws of constant static and
kinetic friction.  As long as $\dot x<v_0$ the system behaves like an
undamped harmonic oscillator with the equilibrium position shifted by
the amount of $F_K/\kappa$. Thus there are infinitely many
oscillatory solutions.  Below we will see that some of them may
survive in the case of velocity dependent kinetic friction. The
equilibrium position of the block is $x=F_K(v_0)/\kappa$. It is
called the {\em continuously sliding state\/}.

Every initial state which would lead to an oscillation with a
velocity amplitude exceeding $v_0$ leads in a {\em finite\/} time to 
stick-slip motion. Independent of the initial condition the slips
always start with $x=F_S/\kappa$ and $\dot x=v_0$. Thus the
stick-slip motion defines an attractive limit cycle in phase space.
This is not in contradiction with the fact that the system behaves
otherwise like an undamped harmonic oscillator. The reason for that
is that a finite bounded volume in phase space is contracted onto a
line if it hits that part in phase space which is defined by
(\ref{eqm.stick}).  Stick-slip motion requires a kinetic friction
$F_K$ which is strictly less than the static one. Usually the
sticking time $t_{stick}=2(F_S-F_K)/(\kappa v_0)$ is much larger than
the slipping time $t_{slip}=2\bigl(\pi-\arctan[
(F_S-F_K)v_0^{-1}(\kappa M)^{-1/2}]) \sqrt{M/\kappa}$. The maximum
amplitude of the stick-slip oscillation [i.e., $\max_tx(t)$] is a
monotonically increasing function of $v_0$ which starts at
$F_S/\kappa$ for $v_0=0$. This is also true for a velocity-dependent
kinetic friction force.

The unmodified Coulomb's law lead to a coexistence of the
continuously sliding state and stick-slip motion for any value of the
sliding velocity $v_0$. In the more general case of a velocity
dependent kinetic friction this bistability still occurs but in a
restricted range of $v_0$. Especially there will be always a
critical velocity $v_c$ above which stick-slip motion disappears.
This is an everyday experience: Squeaking of doors can be avoided by
moving them faster.

In order to be more quantitative we solve the equation of motion for
a linear dependence of $F_K$ on $v$, i.e., $F_K(v)=F_{K0}+\gamma v$,
with $\gamma>0$.  Eq.~(\ref{eqm.slip}) becomes the equation of a
damped harmonic oscillator which can be easily solved. Instead of a
continuous family of oscillatory solutions we have an attractive
continuously sliding state. Stick-slip motion disappears if the
trajectory with $x(0)=F_S/\kappa$ and $\dot x(0)=v_0$ never sticks
for $t>0$. The critical velocity $v_0=v_c$ is defined by
$x(t_{slip})=F_{K0}/\kappa$ and $\dot x(t_{slip})=v_0$.  It leads to
two nonlinear algebraic equations for $t_{slip}$ and $v_c$. For
$\gamma\ll \sqrt{M\kappa}$ the solution can be given approximately:
\begin{equation}
  v_c=\frac{F_S-F_{K0}}{\sqrt{2\pi\gamma\sqrt{\kappa M}}}
      +{\cal O}(\sqrt{\gamma}).
  \label{vc}
\end{equation}
The critical velocity $v_c$ plays an important role in the discussion
of the nature of stick-slip motion, because its measurement tells us
indirectly something about the mechanisms of dry friction (see the
discussion in \cite{ber.96}).

Next we discuss a general nonmonotonic $F_K(v)$ like the examples
shown in Fig.~\ref{f.fk}. The static friction $F_S$ is still assumed
to be constant. The continuously sliding state exists for all values
of $v_0$ but it is stable only if $F_K'\equiv dF_K(v_0)/dv_0>0$. At
an extremum of $F_K(v)$ the stability changes and a Hopf bifurcation
occurs. Near the extremum and for small deviations from the
continuously sliding state the dynamics of 
\begin{equation}
  x(t)-\frac{F_K(v_0)}{\kappa}=A(t)e^{i\sqrt{\kappa/M}t}+\mbox{c.c.}
  \label{x.A}
\end{equation}
is governed by the amplitude equation (normal form) \cite{kev.81}
\begin{equation}
  \frac{dA}{dt}=-\frac{F_K'}{2M}A-\left(\frac{\kappa F_K'''}{4M^2}
   +i\left(\frac{F_K''}{M}\right)^2\sqrt{\frac{\kappa}{M}}\right)
   |A|^2A.
  \label{Amp.eq}
\end{equation}
If the third derivative of the kinetic friction at an extremum is
positive, the Hopf bifurcation is supercritical, and in addition to 
the well-known attractors mentioned above, another type of attractor
appears. I call it the {\em oscillatory sliding state\/}. It is a
limit cycle where the maximum velocity remains always less than
$v_0$.  Thus the block never sticks. Its frequency is roughly given
by the harmonic oscillator of the left-hand side of (\ref{eqm.slip}).
The second derivative of the kinetic friction is responsible for
nonlinear frequency detuning. Note that the frequency of the
stick-slip oscillator is usually much smaller than the frequency of
the oscillatory sliding state. This oscillatory state is similar to
the limit cycle of Rayleigh's equation $\ddot u+\epsilon(\dot
u^3-\dot u)+u=0$ \cite{kev.81}, in fact, Rayleigh's equation is a special
case of (\ref{eqm.slip}). Depending on the kinetic friction, several
stable and unstable limit cycles may exist. By varying $v_0$ they are
created or destroyed in pairs due to saddle-node bifurcations. 

It should be noted that the Hopf bifurcation described by
(\ref{Amp.eq}) is not related to the Hopf bifurcation observed by
Heslot {\em et al.\/} \cite{hes.94} which occurs in a regime (called
creeping regime) where (\ref{eqm}) is not applicable (see also the
discussion below about the validity of dry friction laws).

An oscillatory sliding state exists only if its maximum velocity is
smaller than the sliding velocity $v_0$ because of the sticking
condition (\ref{eqm.stick}). How does the interplay of the
oscillatory sliding states and the sticking condition leads to
stick-slip motion? In order to answer this question we calculate the
backward trajectory of the point $\lim_{\epsilon\to 0}(F_K(0)/\kappa,
v_0-\epsilon)$ in accordance with (\ref{eqm.slip}). Three
qualitatively different backward trajectories are possible:
\begin{enumerate}
\item The backward trajectory hits the sticking condition.  Together
they define a bounded set of initial conditions leading to
non-sticking trajectories. I call the boundary of this set the
special stick-slip boundary; it is not a possible trajectory but it
separates between the basins of attraction of the stick-slip
oscillator and the non-stick-slip attractors.
\item The backward trajectory spirals inwards towards an unstable
oscillatory or continuously sliding state. Again all initial states
outside these repelling states are attracted by a stick-slip limit
cycle. 
\item The backward trajectory spirals outward towards infinity, and 
stick-slip motion is impossible. 
\end{enumerate}

\begin{figure}
\epsfxsize=80mm\epsffile{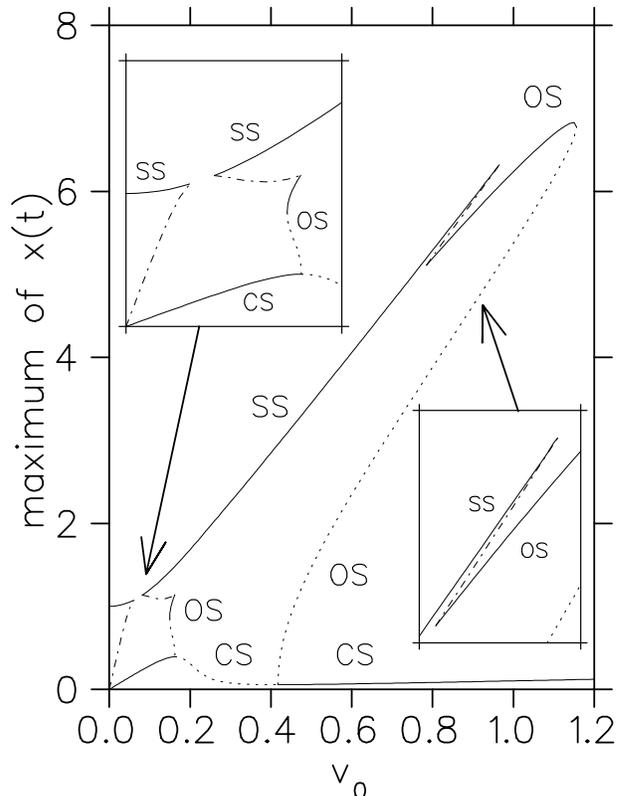}
\caption[fbif]{\protect\label{f.bif}Typical bifurcation scenarios for
a particular kinetic friction force $F_K(v)$ of type shown in
Fig.~\ref{f.fk}(b). The following function has been chosen
$F_K(v)=v[\gamma_1+\gamma_2+(\gamma_2-\gamma_1)(v-\tilde{v})/ \sqrt{(
\Delta v)^2+(v-\tilde{v})^2}]/2$, with $\gamma_1=3$, $\gamma_2=0.1$,
$\tilde{v}=0.2$, and $\Delta v=0.05$. The results are obtained by
numerical integration of the equation of motion (\ref{eqm}). The
other parameters are $F_S=1$, $M=40$, and $\kappa=1$. Solid (dotted)
lines indicate stable (unstable) continuously sliding states (CS),
oscillatory sliding states (OS), or stick-slip motions (SS). The
dashed-dotted line indicates the special stick-slip boundary.}
\end{figure}

Two types of local bifurcations are possible:  If the backward
trajectory changes from case~1 to case~3 the stick-slip limit cycle
annihilates with the special stick-slip boundary. For changes from
case~1 to case~2 the special stick-slip boundary is either replaced
by an unstable continuous or oscillatory sliding state or it
annihilates with a stable continuous or oscillatory sliding state. A
change from case~2 to case~3 is not possible. Figure~\ref{f.bif}
shows for a particular choice of $F_K(v)$ both types of bifurcations.
Here the first bifurcation type occurs at $v_0\approx 0.059$, 0.082,
and 0.966. The second type occurs at $v_0\approx 0.162$ and 0.785.
This example shows that for increasing $v_0$ stick-slip motion can
disappear and reappear again.

Besides of the well-known bistability between stick-slip motion and
continuous sliding \cite{hes.94}, multistability between one
continuously sliding state, several oscillatory sliding states, and
one stick-slip oscillator is possible (see Fig.~\ref{f.bif}). 
Eventually for large sliding velocities all attractors except that of
the continuously sliding state will disappear because the kinetic
friction has to be an increasing function for sufficiently large
sliding velocities. 

The strongly overdamped limit (i.e., $|dF_K(v)/dv|\gg \sqrt{\kappa
M}$ for any $v$ except in tiny intervals around the extrema) leads to
a separation of time scales: From an arbitrary point $(x,\dot x)$ in
phase space with $\dot x<v_0$ the system moves very quickly into the
point $(x,v)$ where $v$ is a solution of $\kappa x= F_K(v_0-v)$ with
$F_K'(v_0-v)>0$. Points on the curve $\kappa x= F_K(v_0-\dot x)$ with
$F_K'<0$ are unstable. They separate basins of attraction of
different solutions $v$.  After the fast motion has decayed the
system moves slowly on the curve $\kappa x= F_K(v_0-\dot x)$. The
direction is determined by the sign of $\dot x$. It either reaches a
stable continuously sliding state, or, near an extremum of $F_K$, it
jumps suddenly to another branch of the curve or to the sticking
condition.  For kinetic friction laws of the form shown in
Fig.~\ref{f.fk}(b) with $v_0$ between the two extrema, we get an
oscillatory sliding state. It is a relaxation oscillation which may
be difficult to distinguish from a stick-slip oscillation. In the
case of a friction law with a single minimum at $v=v_m$ as shown in
Fig.~\ref{f.fk}(a) we get stick-slip motion for $v_0<v_m$
\cite{per.94}. In the strongly overdamped limit any multistability
disappears except near the extrema of $F_K(v)$. The experiments of
Yoshizawa and Israelachvili \cite{yos.93} are consistent with the
assumption that the system is in a strongly overdamped limit with a
friction law as shown in Fig.~\ref{f.fk}(a) \cite{per.94}.

In order to discuss the influence of a stick-time dependent static
friction on the stick-slip behavior we define a stick-slip map
$x_{n+1}=T(x_n)$, where $x_n$ is the position of the block just
before slipping. For constant static friction the map reads
$T(x)=F_S/\kappa$. The position just at the time of the slip-to-stick
transition is defined by $x_n^s$. It is a function of $x_n$, i.e.,
$x_n^s=g(x_n)$, where $g$ is usually a monotonically decreasing
function.  The sticking time $t_n^{stick}$ is the smallest positive
solution of 
\begin{equation}
  \frac{F_S(t_n^{stick})}{\kappa}=x_n^s+v_0t_n^{stick}.
  \label{t.stick}
\end{equation}
This defines a function $t_n^{stick}=h(x_n^s)$ which is a
monotonically decreasing function due to $F_S'>0$. Thus the
stick-slip map is given by $T(x)=F_S(h(g(x)))/\kappa$. If the map has
one fixed point, then stick-slip motion exists. 

For $F_K=const=F_S(0)$, stick-slip motion disappears if
$v_0>v_c=\sup_{t>0}2[F_S(t)-F_S(0)]/(\kappa t)$. For a non-convex
$F_S(t)$ the supremum occurs at a non-zero value of the sticking time
leading to a saddle-node bifurcation of a stable and an unstable
fixed point of the stick-slip map. At $v_0=v_c$ the stick-slip motion
has a finite amplitude, contrary to the case of a convex $F_S(t)$
\cite{per.95}. Because $T$ is a monotonically increasing function,
limit cycles or even chaos are not possible. If the slip-to-stick
transition does not happen at the first time when $\dot x$ becomes
equal to $v_0$ [because of $|x_n^s|>F_S(0)/\kappa$] chaotic motion
may occur \cite{vet.90}. In this case we get a non-monotonic $T$ due
to a non-monotonic $g$. Such over-shooting is only possible if
$F_S(\infty)/F_S(0)$ becomes relatively large. For example, for a
constant kinetic friction over-shooting occurs if
$F_S(\infty)/F_S(0)>1+F_K(0)/F_S(0)$. For most realistic systems this
condition is not satisfied.  Note that the possibility of chaos is not
in contradiction with the fact that the equation of motion
(\ref{eqm}) with constant $F_S$ can not show chaotic motion. But the
retardation of $F_S$ turns (\ref{eqm}) into a kind of
differential-delay equation.

Using phenomenological dry friction means that we treat dry friction
as an element in a mechanical circuit with some nonlinear
velocity-force characteristic like, say, a diode in an electrical
circuit. This treatment is justified as long as the macroscopic time
scales are much larger than any time scale of the internal degrees of
freedom of the interacting solid surfaces. But there is one internal
time scale which diverges if the relative velocity between the
surfaces goes to zero: It is given by the ratio of a characteristic
lateral length scale of the surface and the relative sliding
velocity. Thus any kinetic friction law $F_K(v)$ becomes invalid if
\begin{equation}
  v\lesssim\frac{\mbox{microscopic length scale}}
   {\mbox{macroscopic time scale}}.
  \label{invalid} 
\end{equation} 
The characteristic length scale 
ranges from several micrometers to several meters. It may be the size
of the asperities, the size of the contact of the asperities, the
correlation length of surface roughness, or an elastic correlation
length. This limitation of any dry friction law does not concern
oscillatory sliding states and continuously sliding states, as long
as their relative sliding velocity stays always much larger than the
critical velocity (\ref{invalid}). But the transition between
sticking and sliding in a stick-slip motion may be strongly affected
by the fact that just after stick-to-slip transitions and just before
slip-to-stick transitions, details of the interface dynamics become
important. One may expect that the importance of these details
increases when the maximum slipping velocity decreases. For example,
Heslot {\em et al.\/} \cite{hes.94} found experimentally a completely
different behavior when the maximum relative sliding velocity during
a slip was below the critical value (\ref{invalid}).

In this paper the nonlinear dynamics of a harmonic oscillator sliding
over a solid surface has been discussed under the assumption that dry
friction can be described by a velocity dependent kinetic friction
and a sticking-time dependent static friction. Besides of the
well-known continuously sliding state and the stick-slip oscillator,
an oscillatory sliding state without sticking has been found. All
typical bifurcation scenarios of these states are shown in
Fig.~\ref{f.bif}. 

\acknowledgments
I gratefully acknowledge H. Thomas for his critical reading of the
manuscript.  This work was supported by the Swiss National Science
Foundation.


\begin{references}
\bibitem{bow.54}
F.~P. Bowden and D. Tabor, {\it Friction and Lubrication} (Oxford
University Press, 1954).

\bibitem{rab.65}
E. Rabinowicz, {\it Friction and Wear of Materials} (Wiley \&
Sons, New York, 1965).

\bibitem{hes.94}
F. Heslot, T. Baumberger, B. Perrin, B. Caroli, and C. Caroli, Phys.
Rev. E {\bf 49}, 4973 (1994).

\bibitem{bur.67}
R. Burridge and L. Knopoff, Bull. Seismol. Soc. Am. {\bf 57}, 341 
(1967).

\bibitem{bhu.95}
B. Bhushan, J.~N. Israelachvili, and U. Landman, Nature {\bf 374},
607 (1995).

\bibitem{ber.96}
A.~D. Berman, W.~A. Ducker, and J.~N Israelachvili, in {\em The 
physics of sliding friction\/}, B.~N.~J. Persson and E. Tosatti 
(eds.), (Kluwer Academic Publishers, Dordrecht, 1996).

\bibitem{per.94}
B.~N.~J Persson, Phys. Rev. B {\bf 50}, 4771 (1994).

\bibitem{per.95}
B.~N.~J Persson, Phys. Rev. B {\bf 51}, 13568 (1995).

\bibitem{vet.90}
M.~M. Vetyukov, S.~V. Dobroslavskii, and R.~F. Nagaev, Izv. AN SSSR.
Mekhanika Tverdogo Tela {\bf 25}, 23 (1990) [Mech. of Solids {\bf
25}, 22 (1990)].

\bibitem{isr.85}
J.~N. Israelachvili, {\it Intermolecular and Surface Forces}
(Academic Press, London, 1985).

\bibitem{mat.87}
C.~M. Mate, G.~M. McClelland, R. Erlandsson, and S. Chiang, Phys.
Rev. Lett. {\bf 59}, 1942 (1987).

\bibitem{jae.96}
For an overview and more references on the physics of granular
materials see H.~M. Jaeger, S.~R. Nagel,
and R.~P. Behringer, Physics Today {\bf 49}, No.~4, 32 (1996).

\bibitem{kev.81}
J. Kevorkian and J.~D. Cole, {\it Perturbation Methods in Applied
Mathematics} (Springer, New York, 1981).

\bibitem{yos.93}
H. Yoshizawa and J. Israelachvili, J. Phys. Chem. {\bf 97}, 11300
(1993).

\end{references}
\end{document}